\documentclass[journal=ancac3,manuscript=article]{achemso}
\setkeys{acs}{
	keywords     = true,
	maxauthors  = 20
}

\usepackage[version=3]{mhchem} 
\usepackage[T1]{fontenc}       
\usepackage{color}

\title{Vapor Condensed and Supercooled Glassy Nanoclusters}

\author{Weikai Qi}

\author{Richard K. Bowles}
\email{richard.bowles@usask.ca}

\affiliation{Department of Chemistry, University of Saskatchewan, Saskatoon, S7N 5C9, Canada }


\begin{document}


\begin{abstract}

We use molecular simulation to study the structural and dynamic properties of glassy nanoclusters formed both through the direct condensation of the vapor below the glass transition temperature, without the presence of a substrate, and \textit{via} the slow supercooling of unsupported liquid nanodroplets. An analysis of local structure using Voronoi polyhedra shows that the energetic stability of the clusters is characterized by a large, increasing fraction of bicapped square antiprism motifs. We also show that nanoclusters with similar inherent structure energies are structurally similar, independent of their history, which suggests the supercooled clusters access the same low energy regions of the potential energy landscape as the vapor condensed clusters despite their different methods of formation. By measuring the intermediate scattering function at different radii from the cluster center, we find that the relaxation dynamics of the clusters are inhomogeneous, with the core becoming glassy above the glass transition temperature while the surface remains mobile at low temperatures. This helps the clusters sample the highly stable, low energy structures on the potential energy surface. Our work suggests the nanocluster systems are structurally more stable than the ultra-stable glassy thin films, formed through vapor deposition onto a cold substrate, but the nanoclusters do not exhibit the superheating effects characteristic of the ultra-stable glass states.
\end{abstract}


 \vspace{12pt}
This document is the Accepted Manuscript version of a Published Work that appeared in final form in ACS Nano, copyright \textcircled{c} American Chemical Society after peer review and technical editing by the publisher. To access the final edited and published work see http://pubs.acs.org/doi/abs/10.1021/acsnano.5b07391.

\vspace{12pt}

Glassy materials are amorphous solids usually formed by rapidly cooling a liquid below its equilibrium freezing temperature to kinetically trap the particles in a liquid-like structure at the glass transition temperature, ($T_g$)~\cite{AustenAngell:1995p415,Debenedetti:2001p1730,Berthier:2011p14519}. They can be formed from a wide variety of materials, including metallic alloys, organic molecular systems, and covalently bonded network forming silicates, and they play essential roles in a range of technologies such as high strength materials, optical fibers and electronic components. Similarly, amorphous nanoparticles are used in a variety of advanced applications in catalytic, optical, magnetic and biological materials~\cite{Hoang:2012cl}. However, despite the importance of these systems, and our considerable ability to control and manipulate the properties of amorphous materials, a genuine understanding of what gives rise to their glassy behavior remains elusive~\cite{Kivelson:2008jg, Ediger:2012fw, Langer:2014ic}.  

Thermodynamic arguments suggest the glass transition is an experimental manifestation of an underlying thermodynamic transition to an ideal glass state that occurs at a much lower temperature known as the Kauzmann temperature~\cite{kauzmann48} ($T_K$). In this approach, the multi-dimensional potential energy landscape~\cite{Goldstein:1969uq} is divided into local basins of attraction consisting of the configurations that map, through a steepest descent or conjugate gradient quench, to a local potential energy minimum. The mechanically stable configuration at the minimum is an inherent structure~\cite{Stillinger:1982jb} of the system and the configurations in its basin of attraction represent local thermal excitations. The thermodynamics and dynamics of the liquid are then described in terms of how the system samples the basins and the saddle points that separate them. At high temperatures, the liquid samples a region of the landscape that is characterized by a large number of high energy inherent structures, but as it cools the system trades entropic stability, measured in terms of the number of accessible inherent structure basins, for energetic stability by sampling rarer, lower energy basins. If $T_K$ could be reached, the liquid is expected reach the bottom of the landscape, corresponding to an ideal glass state, where there are a sub-exponential number of inherent structures basins. The configurational entropy is then sub-extensive and the system avoids the Kauzmann entropy catastrophe as absolute zero is approached. Alternative theories, such as geometric frustration~\cite{Tarjus:2005p5454}, focus on the role of local structures and their ability to prevent the formation of the crystal. The icosahedron is the primary candidate in three dimensions because it is a low energy structure that is unable to tile Euclidean space~\cite{Frank:1952ux,Johsson:1988ve}, but recent studies~\cite{Coslovich:2007bi,PatrickRoyall:2008fz,Malins:2014cx} have begun to analyze a broader range of local polyhedral motifs that can induce geometrical constraints, either due to their intrinsic inability to tile space or as a result of the motifs becoming distorted through compositional effects. In particular, the presence of locally favored polyhedral motifs, such as the bicapped square antiprism, have been connected with dynamically slow domains in the Kob-Andersen (KA) model~\cite{Kob:1995dh} for the $Ni_{80}P_{20}$ binary alloy, which is a classic glass forming model. However, not all local structure theories rely on geometric frustration, and polyhedral ordering associated with crystal-like structure has also been linked to elements of glassy behavior such as dynamic heterogeneity~\cite{Tanaka:2010jo}.

Recent experiments have shown that ultrastable glasses, with low energies and an enhanced kinetic stability, can be formed through physical vapor deposition (PVD) onto a substrate held at a temperature, $T_d$, below the glass transition temperature~\cite{Swallen:2007fh,Singh:2013fm,Lyubimov:2013kl}. At the optimal temperature, $T_d\approx0.8T_g$, the ultra-stable glasses exhibit relaxation times 2-3 orders of magnitude slower than ordinary glasses and have been shown to remain stable to temperatures well above $T_g$ as they are heated. The stability of these PVD films is thought to be derived from the ability of the newly deposited atoms to diffuse around the free surface to find a low energy local structure before they become permanently trapped by the atoms that follow. This is supported by simulations that show an increased presence of stable local polyhedral motifs in more stable glasses~\cite{Singh:2013fm} and a greater mobility of atoms near the free surface. These systems have the potential to offer new insight into glassy behavior because they appear to be significantly more stable than ordinary, supercooled glasses. However, the relationship between the different glasses is not clear. The ultra-stable glasses could simply be more stable extensions of the supercooled glasses~\cite{Chen:2013ja} or they could be thermodynamically and structurally distinct materials with properties that arise out of their unique history~\cite{Angell:2012ky,Bhattacharya:2015dy}.

Our primary goal is to understand the structural and thermodynamic relationship between vapor condensed glasses and their traditional supercooled counterparts created in the absence of a substrate. However, we are also interested in examining how particle formation history may effect the properties of glassy nanoparticles. Glassy dynamics in organic aerosol particles has been shown to have a strong effect on the ability of a particle to nucleate ice in the atmosphere~\cite{Berkemeier:2014df} but little is known about how these aerosols are formed and different particle histories may lead to contrasting properties.
To address these questions, we use  molecular simulation to explore the properties of glassy nanoclusters formed through traditional supercooling (SC) and to compare them to nanoclusters formed \textit{via} a vapor condensation (VC) approach. We study the thermodynamic, structural and dynamic properties of nanoclusters, formed from a binary mixture of $N=600$ Lennard-Jones atoms, with interaction parameters and a composition consistent with the Kob-Andersen model~\cite{Kob:1995dh}, which is a well known bulk glass former and has been used in the study of ultra-stable glassy films. Our supercooled glassy nanoclusters (SCGN) are prepared by cooling the well equilibrium gas state at a high temperature to form a liquid droplet, with cooling rates in the range $\gamma=3.3 \times 10^{-3} - 3.3 \times 10^{-6}$. The vapor condensed glassy nanoclusters (VCGN), are formed through the direct condensation of the vapor into a nanocluster at a specific temperature, $T_d$, using a technique that simulates the vapor condensation process and which is similar to that employed in the study of thin film, ultra-stable glasses~\cite{Singh:2013fm}. The details of the model and simulations methods can be found in the methods section. While our simulation approach is highly idealized, gas aggregation techniques~\cite{Schmidt:2007gd} have been used to study small nanoclusters of ice/water at low temperatures~\cite{Schmidt:2012em}, which suggests a comparison between vapor condensed and supercooled glassy clusters may be experimentally feasible.

\section{Results and Discussion}

\subsection{Nanocluster Energetics}

\begin{figure}
\includegraphics[width=0.9\textwidth]{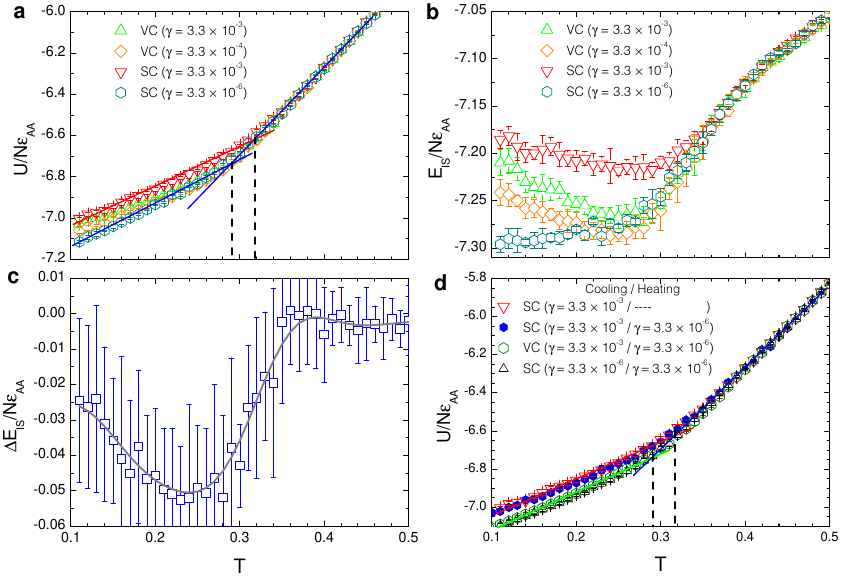}
\caption{\textbf{Nanocluster energetics}. (a) Potential energy per particle for vapor condensation (VCGN) and supercooled (SCGN) glassy nanoclusters prepared at different temperatures with cooling rates as marked.  (b) Inherent-structure energy per particle for VCGN prepared at different deposition temperatures and of SCGN prepared at different deposition or cooling rates as marked. (c) The difference in inherent structure energey $\Delta E_{IS}$ between VCGN and SCGN with a cooling rate of  $3.3 \times 10^{-3}$. The solid line provides a guide to the eye. (d) Potential energy per particle for VCGN and SCGN heated from $T=0.05$ to higher temperatures at a rate of $3.3\times10^{-6}$. Glassy nanoclusters are prepared at deposition or cooling temperature $T = 0.26$ with different cooling rates as marked. For comparison, the potential energy per particle for SCGN formed with a cooling rate of $3.3\times10^{-3}$ is also shown. The energies represent averages taken from ten independent heating runs for each condition. In (a) and (d), the solid lines are linear fitting of the data, and the dashed lines indicate the glasses' transition temperatures.  In all the data, error bars represent the standard deviation of the data for ten independent runs.}
\label{Fig:01}
\end{figure}

Figure 1(a) shows the potential energy per particle, $U/N \epsilon_{AA}$, for the SCGN and the VCGN as a function of temperatures $T$ and $T_d$ respectively. The change in the slope of $U/N\epsilon_{AA}$ indicates the point where the cluster has fallen out of equilibrium on the time scale of our simulations~\cite{Debenedetti:1996fk} and defines the glass transition temperature in our system. The $T_g$ of the supercooled clusters moves from 0.32 down to 0.29 as the cooling rate decreases from $3.3 \times 10^{-3}$ and $3.3 \times 10^{-6}$. These glass transition temperatures are significantly lower than that of the bulk phase KA mixture, for which the mode coupling glass transition temperature, $T_{MCT} = 0.43$. The potential energy of the VCGN formed using $\gamma=3.3 \times 10^{-3}$ is lower than the energy of the corresponding supercooled cluster, but the VCGN formed using $\gamma=3.3 \times 10^{-4}$ and the supercooled cluster formed with $\gamma=3.3 \times 10^{-6}$ exhibit similar energies. Above the highest $T_g$, all the clusters have identical energies, suggesting they are equilibrated liquid drops. 

The inherent structure energies, $E_{IS}$, shown in Fig. 1(b), tell the same story with slower $\gamma$ leading to lower energy structures for both cluster types. The minimum in the $E_{IS}$ of the VCGNs as a function of $T_d$ suggests there is an optimal temperature for their formation. At intermediate temperatures, there is sufficient thermal energy in the system to allow the surface atoms to move and find lower energy sites during the simulation equilibration time, but at very low $T_d$ the atoms remain close to where they deposit on the surface and cannot lower their energy as much. The $E_{IS}$ curve for the supercooled clusters formed with $\gamma=3.3 \times 10^{-3}$ exhibits a similar shape, but with a shallower minimum because the coldest supercooled clusters have evolved through the intermediate temperatures which gives them additional time to age. The data for the slowest cooling rate does not exhibit a minimum, but we would expect one to develop if the equilibration time at each $T$ was sufficiently long.

Figure 1(c) shows the difference in inherent structure energies between vapor condensed and supercooled glassy clusters with $\gamma=3.3 \times 10^{-3}$.  The maximum difference occurs at $T_d \approx 0.25$, or approximately $80 \%$ $T_g$, which is consistent with experiments and simulations of the lowest-energy, vapor deposited film glasses~\cite{Singh:2013fm, Lyubimov:2013kl}. However, we also find that the $E_{IS}$ of the VCGN cooled at $\gamma=3.3 \times 10^{-4}$ have the same energies as the ordinary supercooled glass formed with $\gamma=3.3 \times 10^{-6}$, except at the very lowest $T$. This suggests that the ordinary supercooled glassy nanoclusters can sample the same low energy regions of the inherent structure landscape as the VCGN, but must be cooled at rates 2-3 orders of magnitude more slowly. Such a direct comparison between the glasses formed by the two different methods is not possible in the thin film systems because the supercooled films become trapped in the high energy inherent structures basins. 

Thin film ultra-stable glasses appear to be extremely kinetically stable, a property that is highlighted by their ability to superheat and remain glassy well above the glass transition temperature. To examine this behavior in the nanocluster systems, we plot the potential energy for a number of different glasses as they are heating compared to the cooling curve of the ordinary supercooled nanocluster formed with $\gamma=3.3 \times 10^{-3}$ (Fig. 1(d)). It is immediately obvious that none of the clusters, including those formed through vapor condensation, heat beyond their glass transitions temperatures, which might suggest that the nanoparticle systems do not form ultra-stable glasses. The supercooled nanocluster formed with $\gamma=3.3 \times 10^{-3}$ follows its original cooling curve even though its heating rate is three orders of magnitude slower. In bulk glass samples, this usually leads to some degree of hysteresis. We also compare the melting of two clusters with similar inherent structures ($\delta E_{IS} < 0.01$), but one is prepared by the supercooled method with the cooling rate $3.3 \times 10^{-6}$ and the other is prepared by vapor condensation with the cooling rate $3.3 \times 10^{-3}$. Both clusters are prepared at  $T= 0.26$, and then immediately quenched to $T= 0.05$. They are then heated with a rate $3.3 \times 10^{-6}$. Despite their different histories, the clusters follow very similar heating curves, which suggests that they may be similar in structure and melt in the same way.

\subsection{Nanocluster Structure}
Single component Lennard-Jones nanoclusters freeze to a variety of ordered structures including icosahedra and decahedra~\cite{SaikaVoivod:2010p11422} that contain locally ordered, face-centered-cubic atoms. Using Steinhardt~\cite{Steinhardt:1983uh} based order parameters (see Methods and SI), we find only 1-2 isolated crystalline-type atoms in any of the clusters studied here, which suggests they are amorphous and that the KA mixture remains a good glass former in these nanoscale systems. Figure 2(a) shows the radial density profile, $\rho(R)=N_s(R)/V_s(R)$, where $N_s(R)$ is the number of atoms in a shell with a volume, $V_s(R)$, and a thickness of $0.25\sigma_{AA}$, centered at a radius $R$ from the center of mass of the cluster. The density near the surface is similar for all clusters and varies smoothly, delaying to zero at a radius of about $5.5\sigma_{AA}$, but the core of the rapidly cooled SCGNs ($\gamma=3.3\times 10^{-3}$) have a lower density than the slowly cooled SCGN and the VCGN. The increased fluctuation near the cluster core, signified by an increase in the error bars, reflects the reduced number of atoms in the volume elements associated with the core, but it is also related to the presence of atomic layering caused by the volume exclusion of the atoms near the center. The compositional distribution within the nanoclusters was examined by measuring the probability of finding a small B-type atom at a radius, $R$ (Fig.~2 (b)). While there are compositional fluctuations, $P(N_B)$ is enriched in the core and depleted at the surface, relative to the expected value of 0.2 based on the cluster stoichiometry and the cluster configuration shown in Fig.~2(c) clearly shows there are no B-type atoms on the surface. The enrichment of the core with respect to the B--type atoms increases the number of A--B interactions, which lowers the energy of the cluster compared to the bulk liquid and highlights the importance of surface--core effects giving rise to structural heterogeneity in nanoscale systems. Interestingly, both the SCGN and the VCGN have very similar density and compositional profiles. This is in contrast to the ordinary and vapor deposited thin film glasses formed on a substrate, which exhibit significant differences near the substrate-glass interface.

\begin{figure}
\includegraphics[width=0.5\textwidth]{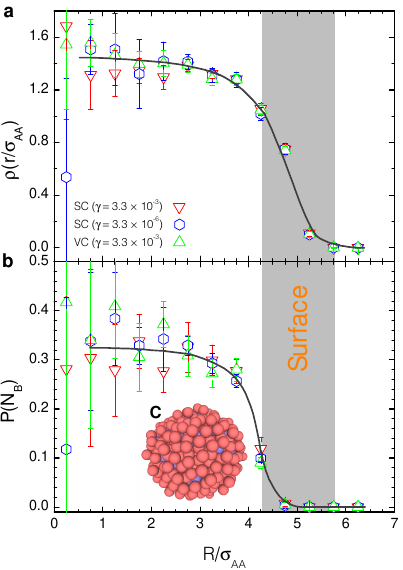}
\caption{\textbf{Radial profiles of nanocluster density and composition.} (a) The density and  (b) the fraction of B particles, as a function of radius from the center of mass of the cluster for the VCGN prepared with  $\gamma=3.3 \times 10^{-3}$ and for SCGN prepared with  $\gamma=3.3 \times 10^{-3}$ and $3.3 \times 10^{-6}$, at temperature $T=0.26$. The solid lines provide a guide to the eye. The grey shading indicates the surface region of the clusters.  (c) A typical configuration of a VCGN at deposition temperature $T_d = 0.26$. The red and blue spheres represent the A and B--type atoms respectively.}
\label{Fig:02}
\end{figure}

The local structure around an atom can be characterized in terms of the geometric properties of its Voronoi cell. In particular, the indices of the Voronoi cell, $\langle n_3, n_4, n_5, n_6 \rangle$ where $n_3$, $n_4$, $n_5$, and $n_6$ are the number of faces shaped as a triangle, quadrangle, pentagon and hexagon, respectively, help identify different types of regular and irregular polyhedra. The sum of the indices also provides the total number of nearest neighbors. Here, we report on the details of the Voronoi cells centered around the B--type atoms, measured in the inherent structures of the clusters and calculated using the  voro++ library~\cite{Rycroft:2009uq}. The number of nearest neighbors of B--type atoms in the glassy clusters is in the range $8 - 13$ (Fig. S3), and has a distribution that is consistent with the properties of the bulk KA glass and with the thin film glasses because all the B--type atoms appear in the core of the nanocluster and there are no surface atoms with a low coordination number. The ordinary glass clusters formed at $\gamma=3.3 \times 10^{-6 }$ and the VCGN formed at $\gamma=3.3 \times 10^{-4}$ are identical within error bars, with only minor differences appearing between the two types of glasses formed at  $\gamma=3.3 \times10^{-3}$.

\begin{figure}
\includegraphics[width=0.95\textwidth]{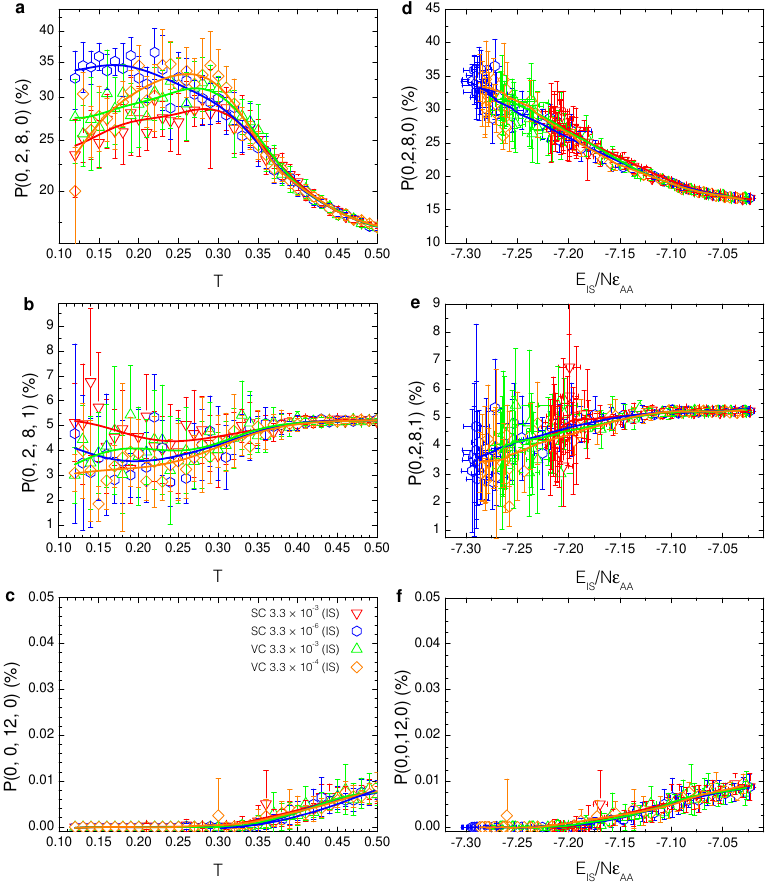}
\caption{ \textbf{Local polyhedral structures as a function of temperature and inherent structure energy.} The fraction of B--type atom Voronoi polyhedra (a)  $\langle 0, 2, 8, 0 \rangle$,  (b) $\langle 0, 2, 8, 1 \rangle$  and (c) $\langle 0, 0, 12, 0 \rangle$ in the inherent structures of the SCGN and VCGN as a function of temperature. Fraction of B-particle Voronoi polyhedra (d) $\langle 0, 2, 8, 0 \rangle$, (e) $\langle 0, 2, 8, 1 \rangle$ and (f) $\langle 0, 0, 12, 0 \rangle$ in the inherent structures of the SCGN and VCGN, formed with different cooling rates, as a function of inherent structure energy per particle. All plots used the same labels as marked in (c). The solid lines represent a smoothing of the data to provide a guide to the eye.}
\label{Fig:03}
\end{figure}

An increased fraction of B--type atoms in a bicapped square antiprism environment, $\langle 0, 2, 8, 0 \rangle$, relative to an ordinary glass, is a key structural characteristic of the thin film ultra-stable glasses, as is an increased fraction of other regular polygons such as the icosahedron, $\langle 0, 0, 12, 0 \rangle$. The $\langle 0, 2, 8, 0 \rangle$ local structure has also been shown to correlate well with the dynamically slow regions in the KA glass~\cite{Malins:2014cx}. Figure 3(a) shows the fraction of $\langle0, 2, 8, 0\rangle$ B--type atoms found in the inherent structures of our nanoclusters as a function of temperature. At $T = 0.5$, approximately $17\%$ of the B--type atoms are in the $\langle0, 2, 8, 0\rangle$ environment, which is close to the value found in the liquid phases of the bulk KA model and the thin film. As the temperature is decreased, the $\langle0, 2, 8, 0\rangle$ fraction continues to increase before eventually reaching a maximum or plateau. The ordinary supercooled clusters formed with $\gamma=3.3 \times10^{-3}$ reach their maximum value of $\sim 28\%$ near $T\approx0.32-0.38$, which is close to the fraction of  $\langle0, 2, 8, 0\rangle$ obtained in the vapor deposited thin film glasses at about the same temperature. The nanoclusters formed with slower cooling rates accumulate an even greater fraction of $\langle0, 2, 8, 0\rangle$, with a maximum of $\sim 35\%$. Figures~\ref{Fig:03}(b) and (c) show that the fraction of $\langle 0, 2, 8, 1 \rangle$ and $\langle 0, 0, 12, 0 \rangle$ decrease with decreasing temperature and are rare in the low energy nanoclusters. In particular, the number of icosahedra in the liquid nanocluster is extremely small and goes to zero above the glass transition temperature. This is surprising because the icosahedron is thought to play a key role in the glassy dynamics of bulk glass formers because they frustrate the formation of the crystal phase. The ability of the nanocluster glasses to accumulate such a significant fraction of $\langle0, 2, 8, 0\rangle$ B--type atoms may result from the compositional changes observed in the cluster core. The bicapped square antiprism polyhedron forms the basis of the Al$_2$Cu crystal which is the lowest energy crystal for the KA model with a 2:1 A-B atom ratio~\cite{Fernandez:2003ux}. This is approximately the same atom ratio observed in the cluster core. However, the crystal structure requires a very specific composition and orientation of the constituent atoms within the local structure in order to tile space and compositional deviations could cause the polyhedron to distort. A recent simulation study of the bulk 2:1 KA model found the total number of $\langle0, 2, 8, 0\rangle$ B--atoms actually decreased by a small amount in the liquid phase relative to the usual 4:1 KA model and the liquid did not freeze to the crystal on simulation timescales~\cite{Crowther:2015cd}. We also do not observe any evidence of crystalization in the nanocluster KA system.

To examine the structural relationship between the glasses formed by vapor condensation and supercooling, we re-plot the data for the fraction of  $\langle 0, 2, 8, 0 \rangle$, $\langle 0, 2, 8, 1 \rangle$ and $\langle 0, 0, 12, 0 \rangle$ local structures as a function of inherent structure energy in Figs. 3(d)-(f), respectively. These clearly show that inherent structures with similar energies exhibit similar structures, which suggests that the two processes sample the same region on the potential inherent structure landscape despite the fact that the clusters are formed in very different ways. A similar result was observed for bulk vapor-deposited and supercooled liquids in two dimensions~\cite{APS1}.

\subsection{Nanocluster Dynamics}

\begin{figure}
\includegraphics[width=0.5\textwidth]{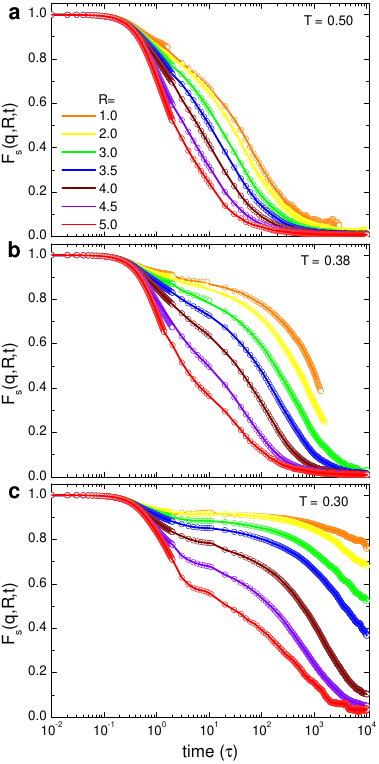}
\caption{\textbf{Inhomogeneous relaxation dynamics as a function of cluster radius}. Shell self intermediate scattering function $F_s(q, R, t)$ at temperatures (a) $T=0.5$, (b) $T=0.38$ and (c) $T=0.30$. The lines provide a guide to the eye.}
\label{Fig:04}
\end{figure}

Our studies suggest that the ordinary supercooled clusters are structurally the same as the VCGN when they are cooled slowly enough. This raises the interesting question of how the supercooled clusters are able to sample the very low energy, stable states accessible to the VCGN. In particular, it has been suggested that surface mobility plays an important role in determining the properties of ultra-stable glasses. To examine mobility as a function of position in the supercooled clusters, we measure a shell self intermediate scattering function, defined as~\cite{Haji-Akbari:2014sf}
\begin{equation}
F_s(q, R, t)=\frac{1}{N_s}\left\langle \sum_{i=1}^{N}\exp [i,\mathbf{q}(|\mathbf{r}_i(t) - \mathbf{r}_i(0)|]D(R(0), R(t), R) \right \rangle
\end{equation}
where $N_s$ is the number of particles in the shell and the function $D(R(0), R(t), R)$ is equal to $1$ when an atom is present in the shell with the distance to the center, $R$, both at the beginning and at the end of the time window, otherwise $D(R(0), R(t), R)$ is equal to $0$. $F_s(q,R,t)$ is evaluated at a wave vector of $q^*\sigma_{AA} = 7.51$, corresponding to the first peak of the static structure factor and the shell thickness is taken to be $0.5\sigma_{AA}$.

Figure 4(a) shows that, at high temperature $T = 0.5$, $F_s(q, R, t)$ decays exponentially with time, indicating that the cluster behaves like a well equilibrated liquid drop. Nevertheless, we see differences in the relaxation times between the surface and the core of the equilibrium liquid drop. Figure 4(b), which shows the $F_s(q, R, t)$ at a temperature $T=0.38$, clearly suggests that structural relaxation within the cluster is not homogeneous and that the core is already deeply supercooled and glassy relative to the outer layers, even though the cluster is still above $T_g$. We see the appearance of a shoulder at longer times, which is characteristic of the slow $\alpha$-relaxation of glass forming supercooled liquids, and this becomes more pronounced with decreasing $T$.  By time the glass transition temperature is reached, the core of the cluster is glassy, but we also see evidence of surface dynamics (Fig. 4(c)). This is consistent with earlier studies~\cite{Hoang:2012fi,Zhang:2012fd} that show the top surface layers of glassy nanoclusters remain dynamically active well below $T_g$, with the presence of liquid-like and solid-like atoms exhibiting heterogeneous dynamics. The inhomogeneous relaxation observed here may help the supercooled clusters sample the very low energy inherent structure basins in a way that is analogous to the free surface dynamics mechanism proposed for the relaxation in the vapor deposited glasses by allowing the atoms away from the core to continue to explore configuration space while an expanding core becomes kinetically trapped as the cluster is cooled.

\section{Conclusion}

Vapor deposition onto a cold substrate slowly introduces material onto a free surface and allows the atoms or molecules time to search out low energy environments before becoming kinetically trapped. The resulting glasses have a large concentration of favored local structures, forming low energy configurations in the potential energy landscape, and they are very kinetically stable relative to ordinary glasses formed through supercooling. The work presented here shows that vapor condensation leads to the formation of glassy nanoclusters that also accumulate a significant degree of locally ordered structure in the form of Voronoi polyhedra with a bicapped square antiprism arrangement.  We find that VCGN exhibit relaxation times that are 2-3 orders of magnitude longer than that of ordinary glass nanoclusters formed by supercooling at the same cooling rate, which is consistent with earlier predictions, but we also show that supercooling is able to produce glass nanoclusters with low energy inherent structures and low glass transitions temperatures that are structurally the same as those formed through vapor condensation. The ability of supercooled nanoclusters to sample the low energy structures in the inherent structure landscape could result from inhomogeneous dynamics within the cluster and the presence of surface mobility. However, despite the structural and energetic stability of our nanoclusters, it is not immediately clear that they are ultrastable in the same sense as the vapor deposited films because they do not exhibit the kinetic stability associated with superheating beyond the glass transition temperature. This may suggest some aspects of the enhanced kinetic stability of the ultrastable thin films arise from the structural anisotropies induced by the presence of the substrate. Haji-Akbari and Debenedetti~\cite{Haji-Akbari:2014sf} recently studied the structural and kinetic effects of substrate interactions in thin films and found that weakly attractive surfaces were able to promote particle mobility near the substrate, relative to the bulk of the film, while strongly attractive surfaces slowed particle mobility. They were also able to correlate these effects to oscillations in the density profile and lateral stress near the substrate surface. The clusters studied here do show some level of density fluctuation in the core, caused by the volume exclusion of the atoms near the center, but the density difference between the core and surface is probably the main cause of the different dynamics observed as a function of cluster radius. Cluster size may also be a key factor in the absence of any superheating. The core of the cluster is never very far from the surface which may help with melting in both the VCGN and SCGN systems. If this is the case, then differences between the two types of clusters might become apparent as the nanoclusters become larger, making it possible to use cluster size to probe the relationships between dynamics, structures and their associated length scales.

The inhomogeneous nature of the dynamics in our clusters also raises interesting questions regarding the definition of the glass transition temperature in these systems. Experimentally, $T_g$ is often defined as the temperature at which the shear viscosity reaches $10^{13}$ poise, and for molecular systems, this corresponds to a relaxation time in the order of 100 seconds. Such long times are not accessible in computer simulations and the cooling rates employed in simulation are orders of magnitude faster than those used in experiment. As a result, model supercooled liquids fall out of equilibrium on the timescale of a simulation at much higher temperatures than observed in experiment. Royall et al~\cite{Royall:2015ii} made an estimate of $T_g$ for the KA model on the experimental time scales by extrapolating the $\alpha$--relaxation time, obtained from the intermediate scattering function, to lower temperatures using a fit to the Vogel--Fulcher--Tammann equation~\cite{vogel.1921,Fulcher:1925fk,Tammann:1926uq}. They found $T_g=0.357$ and the ideal glass temperature, where the relaxation time diverges, $T_0=0.325$. For the 2:1 component KA model $T_0=0.336$~\cite{Crowther:2015cd}. We identify the glass transition temperature as the $T$ where the potential energy curve in Fig.~\ref{Fig:01}(a) exhibits a change in slope, which is characteristic of the system falling out of equilibrium, and find that $T_g$ occurs in the range of 0.32-0.29, depending on the cooling rate. These glass temperatures are close to the $T_0$ values of the bulk systems and well below the extrapolated $T_g$ values. Some of the difference might be accounted for by the choice of definition, \textit{e.g.} relaxation time \textit{versus} energy. However, if relaxation time is to be used to define the $T_g$, which time should we choose? It is clear from our work that the surface region is still able to relax on the timescale of the simulation, even at $T=0.3$, which must also help the core continue to relax over time. On the other hand, it is not obvious how much of the cluster must fall out of equilibrium to cause the potential energy curve to exhibit a change in slope, which leaves open the question as to the best way to define $T_g$ in systems with inhomogeneous dynamics.

 In summary, we have shown, \textit{via} a direct comparison, that the glassy clusters formed through vapor condensation are structurally and thermodynamically the same as those formed through traditional supercooling.  Our work also suggests that it is a combination of the compositional heterogeneity of the clusters and the enhanced mobility of atoms towards the cluster surface that helps the supercooled clusters reach the stable, low energy states on the potential energy surface through the accumulation of a large number of favored local structures. Given that the properties of nano-sized materials are highly sensitive to their structure, the extreme stability of the glassy clusters could have interesting implications for a variety of nanocluster applications, such as catalysis, and the behavior of glassy nanoclusters in the atmosphere. In all, this suggests that future studies on the behaviour of glassy nanoclusters may provide insights into both the  fundamental nature of glasses and important practical applications of nanoclusters.

\section{Methods}
We study binary mixture clusters containing $N=600$ atoms that interact \textit{via} the Lennard-Jones (LJ) potential,
\begin{equation}
U_{\alpha\beta}(r)  = 4\epsilon_{\alpha\beta}\left[\left(\frac{\sigma_{\alpha\beta}}{r}\right)^{12}-\left(\frac{\sigma_{\alpha\beta}}{r}\right)^6\right],
\end{equation}
and have a composition consistent with the bulk glass forming Kob-Andersen model~\cite{Kob:1995dh}.  $80\%$ of the atoms are type A and $20\%$ are type B. The LJ parameters are $\sigma_{AA} = 1.0$, $\sigma_{BB} = 0.88$, $\sigma_{AB} = 0.8$, $\epsilon_{AA} = 1.0$, $\epsilon_{BB} = 0.5$ and $\epsilon_{AB} = 1.5$. The mass of atoms is set to $m = 1.0$. We use reduced units with respect to $\sigma_{AA}$ and $\epsilon_{AA}/k_B$, where $k_B$ is the Boltzmann constant. The potential is cut off at $r_c^{i,j} = 2.5 \sigma_{i,j}$, where $i, j \in A, B$. 

All our molecular dynamics simulations are performed in the canonical ($NVT$) ensemble using the velocity Verlet algorithm with a time step $dt =0.003$ and the unit of time given by $\sqrt{\sigma_{AA}^2 m/ \epsilon_{AA}}$. The temperature is kept constant by velocity rescaling and the unit of temperature is given by $k_B/\epsilon_{AA}$. The supercooled glassy nanoclusters (SCGN) are prepared by cooling the dilute gas with density $\rho_g = 0.15$ which is initially equilibrated for $10^5$ time steps at temperature $T^* = 0.8$. The system is then cooled and condensed into a cluster with cooling rates in the range $3.3 \times 10^{-3} - 3.3 \times 10^{-6}$. 
At each $T$, a configuration of the cluster is equilibrated  for $4\times 10^5$ time steps before configurations are selected for quenching to their inherent structure every 100 time steps over the next $2\times 10^5$ time steps. Inherent structure quenches are performed using the FIRE algorithm~\cite{Bitzek:2006fk}. The results reflect averages and  errors bars measuring the standard deviation obtained from ten independent runs starting from $T=0.8$ gas state.

The vapor condensed glassy nanoclusters (VCGN) are obtained using a procedure that closely follows the vapor deposition method used to form ultrastable thin film glasses~\cite{Singh:2013fm, Lyubimov:2013kl} but the substrate is removed in our simulations. The first 10 LJ atoms, with the 8:2 composition, are added to a container of volume $V=4000\sigma_{AA}^3$, with periodic boundaries, and equilibrated at $T_h=1.0$ for $10^5$ time steps. The atoms are then cooled to a temperature  $T_d$, with cooling rates in the range of $\gamma=3.3 \times 10^{-3}-3.3 \times 10^{-4}$. Cooling is performed by rescaling the atom velocities after each time step. In the absence of a substrate the atoms condense to form a cluster which is then equilibrated at $T_d$ for  $10^5$ time steps before the potential energy is minimized using the FIRE algorithm~\cite{Bitzek:2006fk}, to quench the cluster to its local inherent structure. This inherent structure configuration is used as the starting configuration for the next condensation cycle that begins with the addition of another 10 atoms as before. While the newly introduced atoms are cooled, the atoms in the  cluster at the beginning of the cycle are maintained at a $T_d$. The results reflect averages and  errors bars measuring the standard deviation obtained from ten independent runs.

\begin{acknowledgement}

We would like to thank the Sylvia Fedoruk Canadian Centre for Nuclear Innovation and the Natural Sciences and Engineering Research Council (NSERC) of Canada for financial support. Computing resources were provided by WestGrid, Compute Canada and the University of Saskatchewan.

\end{acknowledgement}

\begin{suppinfo}
Supporting information provides details of our crystal order parameters and additional analysis of the local structure surrounding the atoms in the glassy clusters.
\end{suppinfo}


\providecommand{\latin}[1]{#1}
\providecommand*\mcitethebibliography{\thebibliography}
\csname @ifundefined\endcsname{endmcitethebibliography}
  {\let\endmcitethebibliography\endthebibliography}{}

\end{document}